\documentclass[twocolumn,showpacs,amsmath,amssymb,prd,nofootinbib]{revtex4}
\usepackage{epsfig}
\def\sss{\scriptscriptstyle}
\def\^#1{^{\sss #1}}
\def\_#1{_{\sss #1}}
\def\beq{\begin{equation}}
\def\eeqno#1{\label{#1}\end{equation}}

\def\cmss{{\rm cm~s^{-2}}}

\def\pc{{\rm pc}}
\def\msun{M\_{\odot}}
\def\az{a\_{0}}

\def\l0{\ell\_{0}}

\def\rar{\rightarrow}

\def\l{\lambda}

\def\fN{\phi\_N}
\def\gfN{\grad\fN}

\def\r{\rho}

\def\m{\mu}
\def\n{\nu}

\def\A{\mathcal{A}}

\def\SS{\mathcal{S}}

\def\a{\alpha}
\def\b{\beta}

\def\vr{{\bf r}}

\def\vg{{\bf g}}

\def\vgN{\vg\_N}
\def\S{\Sigma}

\def\Sdz{\S^0\_D}
\def\Sbz{\S^0\_B}

\def\SM{\Sigma\_M}

\def\grad{\vec\nabla}
\def\div{\vec \nabla\cdot}
\def\gf{\grad\phi}
\def\fpg{4\pi G}
\def\tpg{2\pi G}

\def\azg{\A_0}

\def\gN{g\_N}
\def\ups{\Upsilon}

\begin{document}
\title{Universal MOND relation between the baryonic and `dynamical' central surface densities of disc galaxies}
\author{Mordehai Milgrom }
\affiliation{Department of Particle Physics and Astrophysics, Weizmann Institute}

\begin{abstract}
I derive a new MOND relation for pure-disc galaxies: The `dynamical' central surface density, $\Sdz$, deduced from the measured velocities, is a {\it universal function} of only the true, `baryonic' central surface density, $\Sbz$: $\Sdz=\SM \SS(\Sbz/\SM)$, where $\SM\equiv\az/\tpg$ is the MOND surface density constant. This surprising result is shown to hold in both existing, nonrelativistic MOND theories. $\SS(y)$ is derived: $\SS(y)=\int_0^y\n(y')dy'$, with $\n(y)$ the interpolating function of the theory. The relation aymptotes to $\Sdz=\Sbz$ for $\Sbz\gg\SM$, and to $\Sdz=(4\SM\Sbz)^{1/2}$ for $\Sbz\ll\SM$. This study was prompted by the recent finding of a correlation between related attributes of disc galaxies by Lelli et al. (2016). The MOND central-surface-densities relation agrees very well with these results.

\end{abstract}
\maketitle

\section{\label{introduction} Introduction}
MOND \cite{milgrom83,milgrom83a} attributes the mass discrepancies in galactic systems not to dark matter but to a departure from the standard dynamics at low accelerations. Its basic tenets are: (i) Galactic systems showing large mass discrepancies are governed by new dynamics that are invariant under space-time scaling $(\vr,t)\rar\l(\vr,t)$. There, Newton's $G$ is replaced by a scale-invariant constant $\azg$. (ii) The boundary between the standard and the scale-invariant regime is marked by the constant $\az=\azg/G$, which is an acceleration. Thus, much below $\az$ -- `the deep-MOND limit' -- dynamics are scale invariant, and much above it standard dynamics are approached.
References \cite{fm12,milgrom14c} are recent reviews of MOND.
\par
In contradistinction from the dark-matter paradigm, MOND decrees, as a law of physics, that baryons determine the full dynamics of a system. So, MOND predicts that various `dynamical' galaxy properties -- deduced from the measured accelerations -- are tightly correlated with `baryonic' properties -- those deduced directly from the distribution of baryonic mass (see, e.g., the recent Ref. \cite{milgrom14}).
An example of such a `MOND law' is the mass-asymptotic-speed relation (MASSR) \cite{milgrom83a}, which is arguably the most famous prediction of an exact, functional relation. It had predicted a specific version of the `baryonic Tully-Fisher relation'.
\par
Here, I deal with a new MOND law: {\it a functional relation} between a `baryonic' and a `dynamical' property of pure disc galaxies. The one is the central surface (baryonic) density of the disc, $\Sbz$, the other is the total, `dynamically' measured central surface density, $\Sdz$. The impetus to look for such a MOND central-surface-densities relation (CSDR) has come from the recent finding of a correlation between two similar quantities in a large sample of disc galaxies in Ref. \cite{lelli16}.
\par
This MOND CSDR is quite different from other MOND laws of galactic dynamics -- such as the MASSR, or the discrepancy-acceleration relation -- and thus broadens the scope of the MOND codex (see Sec. \ref{discussion}).
\par
Preliminary aspects of the MOND CSDR have been discussed in Ref. \cite{milgrom09} in an approximate way, in terms of the mean density and characteristic size of the galaxy. In particular, it was shown that MOND predicts that, for $\Sbz\gg\SM\equiv\az/\tpg$, we have
$\Sdz-\Sbz\approx\SM$, which implies that $\Sdz/\Sbz\approx 1$. For $\Sbz\ll\SM$, MOND predicts that $\Sdz$ scales as $(\SM\Sbz)^{1/2}$, with the coefficient estimated approximately, and being somewhat system dependent.\footnote{This prediction was made contrary to claims, e.g. in Ref. \cite{donato09}, tantamount to $\Sdz$ being independent of $\Sbz$ in this limit.}
These predictions are born out by the findings of Ref. \cite{lelli16}.
\par
Here, I go rather far beyond these preliminary predictions, and show in Sec. \ref{derivation} that for pure discs, the MOND CSDR, $\Sbz$-$\Sdz$ relation is functional, and I derive $\Sdz(\Sbz)$ analytically for the full $\Sbz$ range. The predicted relation is compared with the data of Ref. \cite{lelli16} in Sec. \ref{data}. Section \ref{discussion} is a discussion.

\section{\label{derivation} Derivation of the MOND $\Sbz$-$\Sdz$ relation}
A true, or `baryonic', mass distribution $\r(\vr)$ (of a galaxy, say) produces a Newtonian acceleration field $\vgN(\vr)=-\gfN$ that is determined from the Poisson equation. In modified-gravity formulations of MOND, which I shall use here, the acceleration field, $\vg(\vr)=-\gf$, is determined from another equation.
\par
One may, on occasion, want to interpret the MOND field in Newtonian terms and define the `dynamical' density of the galaxy as that for which $\vg$ is the Newtonian field:
\beq \r\_D\equiv -(\fpg)^{-1}\div\vg.  \eeqno{i}
This density is made of the baryonic contribution plus a phantom density that a Newtonist would attribute to dark matter. $\r\_D$ is determined, or constrained, from the observed dynamics of the system (rotation curves, light bending, velocity dispersions, etc.)
\par
Consider, as a special case, an axisymmetric, thin, disc galaxy, reflection-symmetric about its  ($x-y$) midplane.
The dynamical, face-on, central surface density is
 \beq \Sdz\equiv \int_{-\infty}^{\infty}\r^0\_D(z)dz, \eeqno{ii}
where we work in cylindrical coordinates with the $z$-axis along the axisymmetry axis, and $\r^0\_D(z)\equiv \r\_D(r=0,z)$.
\par
The assumed symmetries will enter through their implications that along the $z$ axis the Newtonian and MOND accelerations have only a $z$ component, and that they are each equal in magnitude and opposite in sign at $z$ and $-z$. I only use, directly, the assumption that the system is `thin' along the $z$ axis; i.e., all the matter along this axis is concentrated in a thin disc at the origin, where the baryonic surface-density is $\Sbz$. This still holds for a disc that flares, or is not otherwise thin, in the outer parts. But overall flatness may also enter some aspects indirectly (see below).

\par
The baryonic $\Sbz$ is related to $\gN^+$, the absolute value of the Newtonian acceleration just outside the disc, at the origin: $\Sbz=(\tpg)^{-1}\gN^+$.
Because we are dealing with a thin disc, the expression for $\Sdz$ in eq. (\ref{ii}) has two contributions: One from the disc (baryonic plus phantom) -- which is given by $\Sdz(disc)=(\tpg)^{-1}g^+$, where $g^+$ is the absolute value of the MOND acceleration just outside the disc, at the origin. The reflection symmetry is invoked here.

 The other contribution is
\beq \Sdz(out)=2\int_{0^+}^{\infty}\r^0\_D~dz=-\frac{1}{\tpg}\int_{0^+}^{\infty} \div\vg~dz,\eeqno{iii}
where the integration is from just outside the disc.

\subsection{\label{QUMOND} Derivation in QUMOND}
In QUMOND (quasilinear MOND) \cite{milgrom10a}, $\vg=-\gf$ is determined from the field equation
\beq\div \vg= \div [\n(|\vgN|/\az)\vgN], \eeqno{qumon}
where the interpolating function $\n(y)$ has the limits $\n(y\ll 1)\approx y^{-1/2}$, which follows from scale invariance, and $\n(y\rar\infty)\rar 1$ for correspondence with Newtonian dynamics.
Applying Gauss's theorem to get the jump condition at the origin, we have $g^+=\n(\gN^+/\az)\gN^+$.
Thus \beq \Sdz(disc)=\SM y_0\n(y_0)  ,\eeqno{iv}
where $y_0\equiv \gN^+/\az=\Sbz/\SM$, and $\SM$ the MOND surface density \cite{milgrom09}:
\beq \SM\equiv \frac{\az}{\tpg}=
138(\az/1.2\times 10^{-8}\cmss)\msun\pc^{-2}. \eeqno{sigmam}
To calculate $\Sdz(out)$, note that along the $z$-axis (outside the disc, where $\div\vgN=0$) we have from eq.(\ref{qumon}), $\div\vg=\vgN\cdot \grad\n=-|\vgN|d\n/dz=-\az y d[\nu(y)]/dz$, where $y(z)=|\vgN(r=0,z)|/\az$.
I assumed that along the $z$-axis $\vgN$ always points to the origin. This is obvious for a thin disc. It is true for a much larger class of galaxies, but not for any `thick disc'.
Substituting in eq.(\ref{iii}), changing variables to $y$, and integrating by parts, one gets
\beq \Sdz(out)=\SM[-y_0\n(y_0)+\int_0^{y_0}\n(y)~dy].\eeqno{v}
The first term cancels the disc contribution from eq.(\ref{iv}), and we are left with the desired result
\beq \Sdz=\SM\SS(\Sbz/\SM);~~~~~\SS(y)\equiv\int_0^{y}\n(y)~dy.\eeqno{vi}
{\it $\Sdz$ is thus a unique function of $\Sbz$, otherwise independent of the structure of the disc.}
\par
This is a surprising result, because the dynamics is not local. The value of $\r\_D$ at a given point, and its distribution along the $z$-axis, depend on the full mass distribution of the galaxy. Yet, the integral that is $\Sdz$ turns out to depend only on $\Sbz$!
\par
The high-$\Sbz$ asymptote (Newtonian limit) is gotten by taking $\n\equiv 1$, which gives
$\Sdz/\Sbz\rar 1$. The opposite asymptote (the deep MOND limit) is gotten from the scale-invariant limit $\n(y)=y^{-1/2}$, and gives:
$\Sdz\rar (4\SM\Sbz)^{1/2}$. In this limit, $\Sdz(disc)$ and $\Sdz(out)$ contribute equally, $(\SM\Sbz)^{1/2}$ each.
\par
The two asymptotes meet at $\Sbz=4\SM$.
\par
Since everywhere $\n(y)\ge max(1,y^{-1/2})$, the MOND value of $\Sdz$ is always above the two asymptotes.
\par
For example, for the limiting form: $\n(y\le 1)=y^{-1/2},~\n(y\ge 1)=1$, we have from eq.(\ref{vi})
$\Sdz= (4\SM\Sbz)^{1/2}$ for $\Sbz\le\SM$, and $\Sdz= \Sbz+\SM$ for $\Sbz\ge\SM$.

For $\n(y)=[1+(1+4y^{-1})^{1/2}]/2$, which is widely used in MOND rotation curve analysis, eq.(\ref{vi}) gives
\beq \SS(y)=y/2+y^{1/2}(1+y/4)^{1/2}+2{\rm sinh}^{-1}(y^{1/2}/2). \eeqno{eta}

\subsection{\label{AQUAL} Derivation in the nonlinear Poisson theory}
In the nonlinear Poisson formulation of MOND \cite{bm84}, the MOND acceleration field is determined by the field equation
\beq\div[\m(|\vg|/\az)\vg]= -\fpg\rho=\div\vgN, \eeqno{poissona}
 with $\m(x\ll 1)\approx x$, $\m(x\rar\infty)\rar 1$.
Defining $y=x\m(x)$, $y$ has to be a monotonic function of $x$ \cite{bm84}; so the relation can be inverted. Define then $\n(y)$, such that $x=y\n(y)$. The interpolating function $\n(y)$ is the equivalent of the above $\n(y)$ in QUMOND, and has the same large- and small-$y$ limits.

The relation between $g^+$ and $\gN^+$ is the same as for QUMOND. Thus, eq.(\ref{iv}) for $\Sdz(disc)$ holds here as well.
To calculate the integrand in eq.(\ref{iii}) for $\Sdz(out)$, note that along the $z$-axis, and outside the disc, we have from the field equation (\ref{poissona}) ($g\equiv|\vg|$)
\beq \m(x)\div\vg -g~d\m[x(z)]/dz=0,   \eeqno{vii}
where $x(z)\equiv g(z)/\az$.
I assumed that along the $z$-axis, $\vg$ always points to the origin (so $g_z=-g$ for $z>0$). This can be proven \cite{milgrom02} for a thin disc, but is also true for a much larger class, but not for any `thick disc'.
Differentiating the relation $y=x\m(x)$ with respect to $y$, we have: $1=x~(d\m/dz)(dz/dy)+(dx/dy)\m$. Noting also that $\m(x)=1/\n(y)$, extracting from these the expression for $\div\vg$ along the $z$ axis, substituting in the integrand, and integrating by parts, one finds
that expression (\ref{vi}) for $\Sdz$ holds here as well.
As a result, all the corollaries listed for QUMOND carry to the nonlinear Poisson theory.
\subsection{generalizations}
What can be said about the general case of an arbitrary MOND theory, arbitrary dynamical and baryonic surface densities, and any galactic system?
Confine ourselves first to a two-parameter family of self-similar systems -- such as all exponential discs, with constant thickness-to-size ratio -- with a baryonic density distribution $\r(\vr)=\r_0\varrho(\vr/h)$, with the same $\varrho$. For this family, any definition of a baryonic surface density -- local, such as our $\Sbz$, or global, such as the total mass over some area -- is given by $\S_0\equiv\r_0 h$ up to a dimensionless constant. If $\S_d$ is some `dynamical' surface density -- such as our $\Sdz$, or $\S^0\_T$ defined below in Sec. \ref{data} -- then in any nonrelativistic MOND theory, we have on dimensional grounds, $\S_d=\SM f(\S_0,h,G,\az)$, where $f$ is a dimensionless function, constructed from dimensionless combinations of its variables. The only such variable is readily seen to be $\S_0/\SM$ (or some function of it). We thus have for the family  $\S_d=\SM \SS_f(\S_0/\SM)$. In the Newtonian limit, where $\az$ is not available, we must have $\SS_f(y)= \a y$; the dimensionless constant $\a$ depends on the family and choice of surface densities (but not the exact theory). This is just the Newtonian result. In the opposite, deep-MOND limit, scale invariance dictates that $\SS_f(y\rar 0)\rar \b y^{1/2} $, or $\S_d=\b(\S_0\SM)^{1/2}$. This is because under space-time scaling, a dynamical surface density, which is derived as $V^2/Gr$, scales as $\l^{-1}$, while $\S_0$, defined as $M/r^2$, scales as $\l^{-2}$ (and the two sides of any DML relation have to scale in the same way). The dimensionless constant $\b$ may depend on the family, the choice of surface densities, and the exact MOND theory. But, since MOND is assumed not to involve large dimensionless constants, $\a$ and $\b$ should be of order unity, if $\S_d$ and $\S_0$ are bulk properties (and, thus, do not introduce small dimensionless quantities in themselves).
\par
Thus for all system families taken together, any MOND theory would predict a strong $\S_d-\S_0$ correlation, with the two universal asymptotes.
\par
The surprising and helpful aspect of our analysis here is that the MOND CSDR for pure discs is universal, and can be derived in the two full-fledged MOND theories practiced at present (they give $\a=1$, $\b=2$).

\section{\label{data} Comparison with the data}
\begin{figure}
\includegraphics[width=0.9\columnwidth]{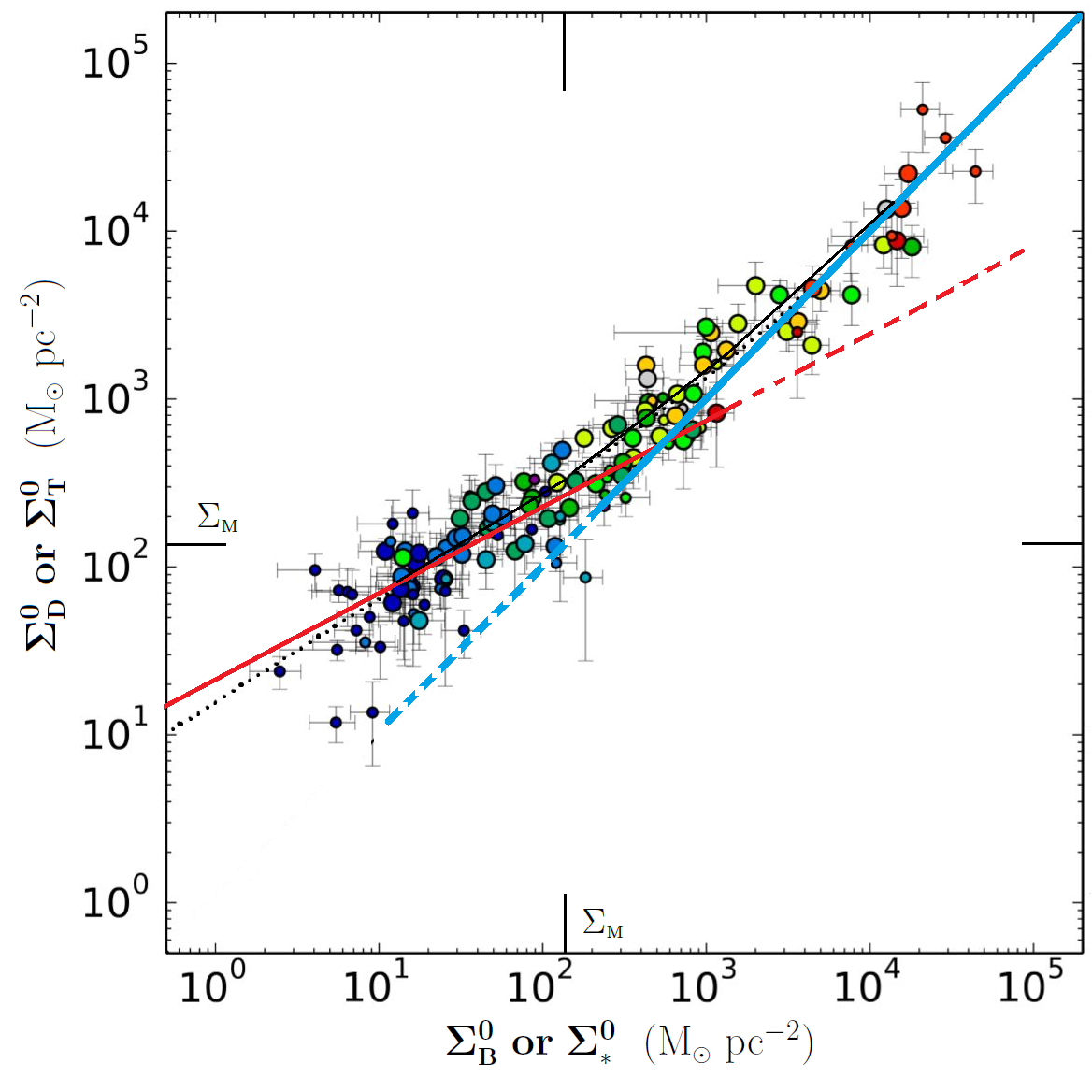}
\caption{\label{fig1}
The $\Sbz$-$\Sdz$ relation. The blue (thicker) line (full and dashed) is the equality line (the Newtonian asymptote of the MOND prediction). The thinner, red line (full and dashed) is the predicted, deep-MOND asymptote. The thinnest, black line is the full MOND relation, eq.(\ref{eta}), for the commonly used $\m(x)=x/(1+x)$. The data points are from Ref. \cite{lelli16}. For the data, the Toomre surface density, $\S^0\_T$, is taken as a proxy for $\Sdz$, and the proxy for $\Sbz$ is the central, stellar surface density, $\S^0_*$.
The dotted line is the best-fit to the data in Ref. \cite{lelli16}, with some 3-parameter formula (not theoretically motivated). No fitting is involved in the MOND curves. The values of the MOND surface density, $\SM$ is marked.
Including the contribution of gas to $\Sbz$ would move some of the low-$\Sbz$ data points somewhat to the right, and correcting for the fact that $0.75<\S^0\_T/\Sdz\le 1$ would move them slightly upward (see comments in subsec. \ref{commdata}).}
\end{figure}

Figure \ref{fig1}  shows the data points of Ref. \cite{lelli16}. For the data, our $\Sbz$ is represented by a proxy: the {\it stellar} central surface mass density, $\S_*^0$, converted from the central surface brightness using a universal mass-to-light ratio in the 3.6 micron photometric band, $\ups=0.5$ solar units.
\par
As a proxy for our $\Sdz$, Ref. \cite{lelli16} used an expression by Toomre \cite{toomre63}: $\S^0\_T=(\tpg)^{-1}\int_0^{\infty}V^2(r)r^{-2}dr$, shown in Ref. \cite{toomre63} to give $\Sdz$ for thin-disc, flat galaxies [$V(r)$ is the rotation curve].\footnote{Other, `dynamical' surface densities one can define are, e.g., $\S^0_n\equiv[\int_0^{\infty}(V^2/\tpg r)^n r^{-1}dr]^{1/n}$. They are sensitive to different regions of the rotation curve for different $n$. $\S^0\_T=\S^0_1$.}

Reference \cite{lelli16} discusses at length why $\S_*^0$, and $\S\_T^0$ -- while not quite the same as $\Sbz$ and $\Sdz$ -- are good proxies. Also shown in Fig. \ref{fig1} is the best fit that Ref. \cite{lelli16} gives for some 3-parameter fit function, not motivated by theory.
\par
The MOND predictions (for $\az=1.2\times 10^{-8}\cmss$) are plotted in Fig. \ref{fig1} as the $\Sbz$-$\Sdz$ plane.
Shown are the asymptotes and the full relation for the widely used choice $\n(y)$ that corresponds to $\m(x)=x/(1+x)$, with $\SS(y)$ from eq. (\ref{eta}).\footnote{I tried several forms of the interpolating function $\n(y)$ and they all give comparable results, hardly distinguishable within the data spread. They all nearly coincide with the deep-MOND asymptote up to $\SM$, and with the `Newtonian' asymptote above $10\SM$; in between, the prediction lies somewhat above the two asymptotes.}

It is seen that the MOND CSDR, which involves no free parameters, agrees very well with the proxy data: the normalization, the asymptotic slopes, the position of the break, and the general behavior.
\subsection{\label{commdata}Some comments on the proxy data}
$\S^0_*$ does not include the contribution from the gas, which is said to be small in most cases. However, some low-$\Sbz$ galaxies are exceptions, and it would have been instructive to include the gas contribution, so as to get the true, baryonic, central surface density.
\par
Departures of $\ups$ from the universal value used by Ref. \cite{lelli16} will contribute to scatter around the predicted relation. And, possible small systematic variations of $\ups$, for example with $\Sbz$, could lead to a small change in the slope.
\par
Even for a pure baryonic disc, $\S^0\_T$, which is exact for thin discs, somewhat underestimates MOND's $\Sdz$, because MOND's $\r\_D$ has a `thick' phantom component [from which $\Sdz(out)$ is made]. For example, it is readily seen that for a spherical system, $\S^0\_T=(1/2)\Sdz$. In the high-$\Sbz$ regime, this should not matter, as the `phantom' component is negligible. In the deep-MOND, low-$\Sbz$ regime, we saw that the `phantom disc' and `phantom halo' contribute equally to $\Sdz$, according to MOND. The disc component is accounted for correctly by the corresponding contribution to $\S^0\_T$. For the `phantom halo' component, $\S^0\_T$ is between $1/2$ and 1 of the contribution to $\Sdz$. In all, we thus expect $\S^0\_T$ to underestimate $\Sdz$ by a factor between $0.75$ and $1$ -- not much of a correction compared with the data spread.
As an example, I calculated the relevant quantities for a deep-MOND, thin, Kuzmin disc, for which the MOND rotation curve is known analytically: $V^2(r)=(M\azg)^{1/2}u^2/(1+u^2)$, with $u=r/h$, and $h$ the Kuzmin scale length \cite{brada95}. I found that $\S^0\_T/\Sdz=\pi/4$, indeed between 0.75 and 1.

\section{\label{discussion}Discussion}
The MOND CSDR differs from other MOND laws (discussed, e.g., in Ref. \cite{milgrom14}) in several important regards: a. It is a relation between a `local' baryonic attribute, $\Sbz$, defined at the center, and a `global' dynamical one: the column dynamical density along the symmetry axis. b. It encompasses the full gamut of accelerations. In comparison, the MASSR relates a global baryonic attribute -- the total mass -- with the asymptotic rotational speeds, and it involves only the deep-MOND regime, and only the outskirts of systems. The MOND `mass-discrepancy-acceleration relation' for disc galaxies, relates the baryonic and dynamical accelerations at the same radius, in the plane of disc galaxies. And, the MOND mass-velocity-dispersion relation (an analog of the virial theorem) relates two global attributes, and holds in the low-acceleration regime.
\par
With good enough data, one can, in principle determine the MOND interpolating function by differentiating the observed $\SS(y)$, from eq. (\ref{vi}). This can also be done in an independent way from the discrepancy-acceleration relation. But this will have to await better data to constrain $\SS(y)$.
\par
The MOND CSDR falls very near the 3-parameter best-fit of Ref. \cite{lelli16}. The scatter around this best fit was shown in Ref. \cite{lelli16} to be $\approx 0.2$ dex, consistent with observational and procedural errors, i.e., with no intrinsic scatter. Furthermore, the residuals do not correlate with various galaxy properties, such as size, gas fraction, or stellar mass.
\par
This is exactly what MOND predicts, while, as also stressed in Ref. \cite{lelli16}, in the dark-matter paradigm, there is no reason why $\Sdz$, which would stand for the central column density of baryon plus dark matter, should be so well correlated with the local $\Sbz$. Especially, the correlation in the low $\Sbz$ region -- where the discrepancy between $\Sdz$ and $\Sbz$ is large -- goes quite against the grain of the cold-dark-matter paradigm: Even with schemes, such as `feedback', `abundance matching', and the like -- put in by hand to save this paradigm from various embarrassments -- one would expect large scatter in any relation between the `dynamical' and baryonic properties, which, to boot, one would expect to depend on galaxy properties (see, e.g., the relevant discussion in Ref. \cite{oman15}).

\end{document}